\pdfoutput=1
\documentclass[sigconf,nonacm]{acmart}

\usepackage{booktabs}
\usepackage{array}
\usepackage{placeins}
\usepackage{enumitem}
\setlist{leftmargin=*, topsep=2pt, itemsep=0pt, parsep=0pt}
\AtBeginDocument{\brokenpenalty=100\relax\widowpenalty=1000\relax\clubpenalty=1000\relax}
\usepackage{xcolor}
\AtBeginDocument{%
  \setlength{\textfloatsep}{6pt plus 2pt minus 2pt}\setlength{\dbltextfloatsep}{6pt plus 2pt minus 2pt}%
  \setlength{\floatsep}{6pt plus 2pt minus 2pt}\setlength{\dblfloatsep}{6pt plus 2pt minus 2pt}%
  \setlength{\intextsep}{6pt plus 2pt minus 2pt}%
  \captionsetup{aboveskip=2pt, belowskip=0pt}}
\DeclareRobustCommand{\R}[1]{\ifcsname r@#1\endcsname\csname r@#1\endcsname\else\textbf{??#1??}\fi}
\expandafter\def\csname r@absent-share-2026\endcsname{25\%}
\expandafter\def\csname r@absent-share-gas-ct-2026\endcsname{30\%}
\expandafter\def\csname r@absent-share-gas-steam-2026\endcsname{74\%}
\expandafter\def\csname r@absent-top1-2026\endcsname{8\%}
\expandafter\def\csname r@cc-fleet-gw\endcsname{23}
\expandafter\def\csname r@cf-base-top1\endcsname{25\%}
\expandafter\def\csname r@cf-both-top1\endcsname{18\%}
\expandafter\def\csname r@cf-both-top5\endcsname{37\%}
\expandafter\def\csname r@cf-drop-share\endcsname{70\%}
\expandafter\def\csname r@cf-sample-top1\endcsname{22\%}
\expandafter\def\csname r@cf-window-top1\endcsname{22\%}
\expandafter\def\csname r@curve-cap-only\endcsname{7\%}
\expandafter\def\csname r@curve-clip-1\endcsname{206\%}
\expandafter\def\csname r@curve-clip-4\endcsname{40\%}
\expandafter\def\csname r@curve-twolayer-1\endcsname{73\%}
\expandafter\def\csname r@curve-twolayer-10\endcsname{11\%}
\expandafter\def\csname r@curve-twolayer-2\endcsname{44\%}
\expandafter\def\csname r@curve-twolayer-30\endcsname{7\%}
\expandafter\def\csname r@curve-twolayer-4\endcsname{25\%}
\expandafter\def\csname r@day-year-eps-1\endcsname{25}
\expandafter\def\csname r@day-year-eps-4\endcsname{164}
\expandafter\def\csname r@delta\endcsname{10^{-6}}
\expandafter\def\csname r@eia923-tx-plants-2025\endcsname{1,021}
\expandafter\def\csname r@eia923-tx-plants-2026\endcsname{530}
\expandafter\def\csname r@ex-base-a\endcsname{39\%}
\expandafter\def\csname r@ex-base-b\endcsname{22\%}
\expandafter\def\csname r@ex-cap\endcsname{760}
\expandafter\def\csname r@ex-ev-rk\endcsname{third}
\expandafter\def\csname r@ex-gap-a\endcsname{1.3\%}
\expandafter\def\csname r@ex-n-fossil\endcsname{80}
\expandafter\def\csname r@ex-ntol\endcsname{8}
\expandafter\def\csname r@ex-nwin\endcsname{70}
\expandafter\def\csname r@ex-off-days\endcsname{150}
\expandafter\def\csname r@ex-off-in-a\endcsname{126}
\expandafter\def\csname r@ex-off-in-b\endcsname{48}
\expandafter\def\csname r@ex-post\endcsname{53\%}
\expandafter\def\csname r@ex-r-a\endcsname{0.92}
\expandafter\def\csname r@ex-r-b\endcsname{0.25}
\expandafter\def\csname r@ex-single-ev\endcsname{12\%}
\expandafter\def\csname r@ex-single-r\endcsname{39\%}
\expandafter\def\csname r@ex-single-static\endcsname{20\%}
\expandafter\def\csname r@ex-static-rk\endcsname{ninth}
\expandafter\def\csname r@fleet-ercot-anon\endcsname{1}
\expandafter\def\csname r@fleet-ercot-cems\endcsname{363}
\expandafter\def\csname r@fleet-ercot-cems-gap\endcsname{21}
\expandafter\def\csname r@fleet-ercot-cems-unique\endcsname{97\%}
\expandafter\def\csname r@fleet-ercot-gens\endcsname{503}
\expandafter\def\csname r@fleet-ercot-unique\endcsname{55\%}
\expandafter\def\csname r@fleet-pjm-anon\endcsname{2}
\expandafter\def\csname r@fleet-pjm-cems\endcsname{729}
\expandafter\def\csname r@fleet-pjm-cems-gap\endcsname{13}
\expandafter\def\csname r@fleet-pjm-cems-unique\endcsname{95\%}
\expandafter\def\csname r@fleet-pjm-gens\endcsname{850}
\expandafter\def\csname r@fleet-pjm-unique\endcsname{50\%}
\expandafter\def\csname r@group-eps-k2\endcsname{22.7}
\expandafter\def\csname r@group-eps-k3\endcsname{38.1}
\expandafter\def\csname r@group-eps-k4\endcsname{56.2}
\expandafter\def\csname r@hub-corr-10\endcsname{0.40}
\expandafter\def\csname r@hub-corr-exact\endcsname{0.59}
\expandafter\def\csname r@inf-day4-2025\endcsname{0.92}
\expandafter\def\csname r@inf-day4-2025-hi\endcsname{0.94}
\expandafter\def\csname r@inf-day4-2025-lo\endcsname{0.87}
\expandafter\def\csname r@inf-month10-2025\endcsname{0.23}
\expandafter\def\csname r@inf-month10-2025-hi\endcsname{0.32}
\expandafter\def\csname r@inf-month10-2025-lo\endcsname{0.13}
\expandafter\def\csname r@inf-month10-2026\endcsname{0.27}
\expandafter\def\csname r@ks-exact-auc-f0\endcsname{0.69}
\expandafter\def\csname r@ks-exact-auc-f99\endcsname{0.96}
\expandafter\def\csname r@ks0-exact-2025\endcsname{0.77}
\expandafter\def\csname r@ks75-exact-2025\endcsname{0.94}
\expandafter\def\csname r@ks90-exact-2025\endcsname{0.99}
\expandafter\def\csname r@laplace-ratio-10\endcsname{10\%}
\expandafter\def\csname r@laplace-ratio-4\endcsname{20\%}
\expandafter\def\csname r@mo-price-10\endcsname{17}
\expandafter\def\csname r@mo-price-10-hi\endcsname{23}
\expandafter\def\csname r@mo-price-10-lo\endcsname{12}
\expandafter\def\csname r@mo-price-4\endcsname{13}
\expandafter\def\csname r@mo-price-4-hi\endcsname{24}
\expandafter\def\csname r@mo-price-4-lo\endcsname{3}
\expandafter\def\csname r@mo-price-true\endcsname{18}
\expandafter\def\csname r@out-exact-2025\endcsname{0.77}
\expandafter\def\csname r@out-exact-2025-hi\endcsname{0.89}
\expandafter\def\csname r@out-exact-2025-lo\endcsname{0.59}
\expandafter\def\csname r@out-exact-2026\endcsname{0.43}
\expandafter\def\csname r@out-fossil-corr-2025\endcsname{0.32}
\expandafter\def\csname r@out-fossil-corr-2026\endcsname{0.26}
\expandafter\def\csname r@out-ren-corr-2025\endcsname{0.90}
\expandafter\def\csname r@out-ren-corr-2026\endcsname{0.81}
\expandafter\def\csname r@out-ren-cov-share-2025\endcsname{95\%}
\expandafter\def\csname r@out-ren-hi-units-2025\endcsname{7}
\expandafter\def\csname r@out-ren-hi-units-2026\endcsname{4}
\expandafter\def\csname r@out-yeardaily10-2025\endcsname{0.16}
\expandafter\def\csname r@out-yeardaily10-2025-hi\endcsname{0.30}
\expandafter\def\csname r@out-yeardaily10-2025-lo\endcsname{0.02}
\expandafter\def\csname r@outage-unit-days\endcsname{56}
\expandafter\def\csname r@outage-unit-group\endcsname{309}
\expandafter\def\csname r@outage-unit-mw\endcsname{600}
\expandafter\def\csname r@plantatk-n-2025\endcsname{130}
\expandafter\def\csname r@plantatk-plant2-10-2025\endcsname{0.30}
\expandafter\def\csname r@plantatk-plant2-10-2025-hi\endcsname{0.46}
\expandafter\def\csname r@plantatk-plant2-10-2025-lo\endcsname{0.12}
\expandafter\def\csname r@plantatk-unit10-2025\endcsname{0.56}
\expandafter\def\csname r@plantatk-unit10-2025-hi\endcsname{0.70}
\expandafter\def\csname r@plantatk-unit10-2025-lo\endcsname{0.36}
\expandafter\def\csname r@plantatk-unit4-2025\endcsname{0.29}
\expandafter\def\csname r@price-twolayer-10\endcsname{3.0}
\expandafter\def\csname r@price-twolayer-4\endcsname{7.4}
\expandafter\def\csname r@qse3-gain-2025\endcsname{0.01}
\expandafter\def\csname r@qse3-gain-2025-hi\endcsname{0.04}
\expandafter\def\csname r@qse3-gain-2025-lo\endcsname{$-$0.04}
\expandafter\def\csname r@qse3-gain-2026\endcsname{$-$0.07}
\expandafter\def\csname r@qse3-hicontrast-2025\endcsname{0\%}
\expandafter\def\csname r@qse3-hicontrast-2026\endcsname{0\%}
\expandafter\def\csname r@qse3-known-2025\endcsname{66\%}
\expandafter\def\csname r@reid-mw-2025\endcsname{32\%}
\expandafter\def\csname r@reid-mw-2026\endcsname{20\%}
\expandafter\def\csname r@reid-n-2025\endcsname{366}
\expandafter\def\csname r@reid-n-2026\endcsname{322}
\expandafter\def\csname r@reid-tech-coal-2025\endcsname{33\%}
\expandafter\def\csname r@reid-tech-gas-cc-2025\endcsname{26\%}
\expandafter\def\csname r@reid-tech-gas-ct-2025\endcsname{32\%}
\expandafter\def\csname r@reid-tech-gas-steam-2025\endcsname{75\%}
\expandafter\def\csname r@reid-tech-solar-2025\endcsname{11\%}
\expandafter\def\csname r@reid-tech-wind-2025\endcsname{6\%}
\expandafter\def\csname r@reid-top5-2025\endcsname{45\%}
\expandafter\def\csname r@reid-top5-2026\endcsname{39\%}
\expandafter\def\csname r@reid-units-2025\endcsname{25\%}
\expandafter\def\csname r@reid-units-2026\endcsname{15\%}
\expandafter\def\csname r@reid-v6-2025\endcsname{18\%}
\expandafter\def\csname r@reid-v9-2025\endcsname{22\%}
\expandafter\def\csname r@reid-v9-2026\endcsname{18\%}
\expandafter\def\csname r@rho-10\endcsname{1.35}
\expandafter\def\csname r@rho-4\endcsname{0.25}
\expandafter\def\csname r@rule1515-cc-days\endcsname{100\%}
\expandafter\def\csname r@rule1515-cc-n\endcsname{65}
\expandafter\def\csname r@rule1515-ct-days\endcsname{100\%}
\expandafter\def\csname r@rule1515-ct-n\endcsname{63}
\expandafter\def\csname r@sigma-month-10\endcsname{1.8}
\expandafter\def\csname r@sigma-month-4\endcsname{4.1}
\expandafter\def\csname r@spike-below50-month\endcsname{34.7}
\expandafter\def\csname r@spike-below50-peak\endcsname{12.6}
\expandafter\def\csname r@spike-head-max\endcsname{36}
\expandafter\def\csname r@spike-head-min\endcsname{5.5}
\expandafter\def\csname r@spike-head-month\endcsname{21}
\expandafter\def\csname r@spike-hub-mean\endcsname{75}
\expandafter\def\csname r@spike-hub-peak\endcsname{1,817}
\expandafter\def\csname r@spike-price-monthcurve\endcsname{39}
\expandafter\def\csname r@spike-price-peakcurve\endcsname{192}
\expandafter\def\csname r@sweep-att-2\endcsname{0.02}
\expandafter\def\csname r@sweep-att-30\endcsname{0.59}
\expandafter\def\csname r@sweep-use-price-2\endcsname{0.29}
\expandafter\def\csname r@sweep-use-price-30\endcsname{0.89}
\expandafter\def\csname r@sweep-use-slope-30\endcsname{0.95}
\expandafter\def\csname r@tab-day1-unit\endcsname{0.62}
\expandafter\def\csname r@tab-month4-unit\endcsname{0.08}
\expandafter\def\csname r@tab-unit-exact\endcsname{1.00}
\expandafter\def\csname r@tab-yeardaily-curve\endcsname{49\%}
\expandafter\def\csname r@tab-yeardaily-unit\endcsname{0.33}
\expandafter\def\csname r@trunc-cost-10\endcsname{0.27}
\expandafter\def\csname r@trunc-eps-10\endcsname{16.4}
\expandafter\def\csname r@trunc-eps-4\endcsname{6.8}
\expandafter\def\csname r@units-per-day-2025\endcsname{1,271}
\expandafter\def\csname r@units-total-2025\endcsname{1,336}
\expandafter\def\csname r@use-fuel-agree-10\endcsname{97\%}
\expandafter\def\csname r@use-headroom-10\endcsname{0.34}
\expandafter\def\csname r@use-headroom-10-pred\endcsname{0.38}
\expandafter\def\csname r@use-mef-10\endcsname{0.21}
\expandafter\def\csname r@use-months\endcsname{33}
\expandafter\def\csname r@use-offered0-10\endcsname{0.66}
\expandafter\def\csname r@use-offered0-10-pred\endcsname{0.68}
\expandafter\def\csname r@use-offered0-4\endcsname{0.37}
\expandafter\def\csname r@use-offered1-10\endcsname{0.13}
\expandafter\def\csname r@use-offered3-10\endcsname{0.17}
\expandafter\def\csname r@use-offered4-10\endcsname{0.04}
\expandafter\def\csname r@use-pred-maxgap\endcsname{0.05}
\expandafter\def\csname r@use-price-10\endcsname{0.66}
\expandafter\def\csname r@use-price-4\endcsname{0.40}
\expandafter\def\csname r@use-slope-10\endcsname{0.64}
\expandafter\def\csname r@use-slope-4\endcsname{0.33}

\newcommand{\eps}{\varepsilon}

\setcopyright{none}
\begin{document}

\title[Your bids are showing]{Your bids are showing: re-identifying power plants in anonymized market data, and a differentially private alternative}

\author{Nikhil Ravi}
\affiliation{\institution{Independent Researcher}\state{Texas}\country{USA}}
\email{thenikhilravi@gmail.com}

\begin{abstract}
ERCOT publishes every generator's hourly offer curve 60 days after the operating day. Market monitors, researchers and traders use these records to study the merit order and check for withholding. PJM publishes the same kind of data with unit names replaced by codes, on the view that masking keeps offers confidential. We test that view on ERCOT data from 2024 to 2026, where names are public and so provide ground truth, by masking the records under PJM's published rules. Public EIA and EPA data match \R{reid-units-2025} of the masked units we can score (\R{reid-mw-2025} of their capacity) to their plant, and \R{reid-top5-2025} fall within the top five guesses. PJM's feed carries the same fields, so we expect masking to fail there too. Publishing exact fuel-group totals instead does not help. An attacker who knows the other units in a group recovers one unit's high-price offers exactly, and an outsider who knows only the target's outage dates reaches a correlation of \R{out-exact-2025}, most of it from a few renewable and hydro units that offer at very high prices (\R{out-fossil-corr-2025} over fossil units). We propose a differentially private release: monthly totals by fuel group and price bin, with Gaussian noise calibrated under zero-concentrated differential privacy so that each unit's whole year of offers is protected. It publishes an unbiased noisy layer and derives a monotone supply curve and shares from it at no extra privacy cost. At $\eps = 10$ per unit-year the market supply curve stays within \R{curve-twolayer-10} of the truth, the implied clearing price within \$\R{price-twolayer-10}/MWh, and the attacker's correlation falls to \R{inf-month10-2025}. Whether an analysis still works depends on one ratio, how much the true quantity varies month to month against the noise, which predicts the accuracy of totals to within \R{use-pred-maxgap}. The clearing price, supply slope and combined-cycle capacity remain usable, but no monthly release, private or exact, can answer questions about single hours or single units.
\end{abstract}
{}

\maketitle

\section{Introduction}
\label{sec:intro}

Every day ERCOT publishes the offer curves, the bids to sell energy at each price, that about \R{units-per-day-2025} generating units submitted to its day-ahead market 60 days earlier, one curve per unit per hour~\cite{ercot60day}. These records are how people outside the market operator see the merit order. Economists use them to measure market power~\cite{hortacsu2008understanding,borenstein2002measuring}, researchers to build supply curves, and traders to anticipate scarcity.

Publishing offers also has a cost. A wholesale electricity market is an auction repeated every hour among a few sellers who know a good deal about each other's costs, and in such a setting information about rivals' offers can help firms settle on high prices without ever communicating, which economists call tacit coordination~\cite{albaek1997government}. Alberta offers direct evidence. From 2000 to 2017 its operator published each hour's offer prices shortly after the hour, without the names of the firms that submitted them. Firms nevertheless responded to particular rivals' offer changes as the report revealed them, which suggests they could tell whose offers were whose~\cite{brown2018information}, and some firms used distinctive offer patterns that appear to have helped coordinate on high prices~\cite{brown2022pricing}. Analysts at the province's market surveillance administrator estimated that publication raised the average price by about 4\%, roughly \$1.1 billion over 2010 to 2015, and the regulator ordered the report discontinued~\cite{olmstead2020offer}.

Operators that publish offers therefore protect them in other ways. ERCOT waits 60 days before it publishes. PJM waits four months and also replaces each unit's name with a code that changes every year~\cite{pjmoffers}, on the reasoning that offers stay confidential as long as nobody can tell whose they are. Other releases replace unit records with totals over many units. Two days after each operating day ERCOT publishes aggregate supply curves for the whole market and for each of its disclosure areas~\cite{ercot2day}, and Nord Pool sells aggregated bidding curves for each bidding zone in Northern and Central Europe~\cite{nordpoolcurves}. Disclosure rules across markets differ mostly in how much they delay, aggregate or mask~\cite{yang2018review}, and in each case the operator picks a point between transparency, which serves oversight, and confidentiality, which limits coordination, without measuring how much of either it obtains.
{}

Neither masking nor aggregation has been tested against an attacker who uses public data. Masking rests on the idea that a code reveals nothing about the unit behind it. A masked unit still offers a capacity and a minimum load and follows a pattern of days on and off, however, and the EIA and EPA publish the capacity, minimum load and hourly operation of every large plant in the country, so the offers themselves can point back to a plant. Aggregation rests on the idea that a unit disappears in a total, an idea that also underlies rules for releasing other energy data. Colorado's 15/15 rule, for example, allows a utility to release a total of customer data if it covers at least 15 customers and no customer accounts for more than 15\% of it~\cite{copuc1515}, a rule that earlier work argued gives no privacy guarantee~\cite{ravi2022differentially}. The idea holds only when the reader knows nothing about the other units in the total, and in a market the owners of those units, or anyone who can model them, know a great deal. Figure~\ref{fig:outage} shows the problem for a \R{outage-unit-mw}~MW combined-cycle unit that was offline for \R{outage-unit-days} days in 2025. Its outages cannot be seen in the exact total of its \R{outage-unit-group}-unit fuel group, but they reappear in full once the other units' contributions are subtracted.
{}

Differential privacy (DP) offers a different kind of promise: a bound on how much a release can change because of any one subject's data, which holds whatever else the attacker knows~\cite{dwork2014algorithmic}. Work on DP in energy has focused on household meter data~\cite{laforet2016individual,wang2020privacy,ravi2022differentially}, on grid models~\cite{fioretto2020differential} and on control~\cite{dvorkin2021differentially,laufer2024privacy}. Market disclosures differ from these settings in three ways that shape the design. The subjects are a few hundred generating units rather than millions of households, so each one carries a large share of any total. Their physical characteristics are public, which gives an attacker strong side information. And the data already has users, from market monitors to traders, whose analyses a private release must keep useful. To our knowledge no prior work has measured what market disclosures reveal or designed a private release for them.
{}

\begin{figure}[tbp]
\centering
\includegraphics{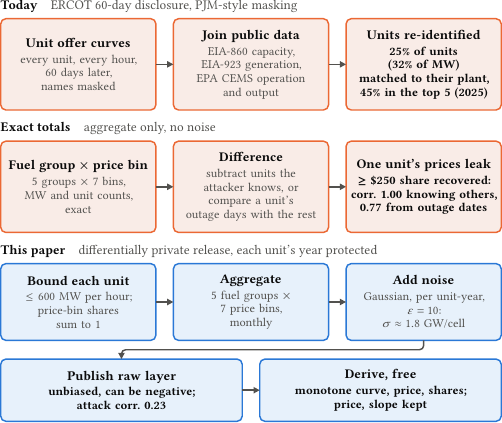}%
\caption{Today's releases and the private release.}
\label{fig:pipeline}
\end{figure}

We use ERCOT as a testbed because it publishes names. That gives us ground truth for every unit, so we can mask the data by PJM's published rules (Section~\ref{sec:pjm-masking}) and score an attacker against the truth. We attack the two release styles in use today, masked unit records and exact totals over many units, with public data only, then design a private release that protects each unit's full year of offers, and test what it keeps for the analyses the data actually serves. Figure~\ref{fig:pipeline} summarizes both halves. We make four contributions, which rest on three kinds of evidence. The privacy guarantee of the release is proved (Theorem~\ref{thm:privacy}). The success of the attacks and the utility of the release are measured on ERCOT data against ground truth, with parameters fixed on 2024 and results reported on 2025 and 2026. That the re-identification attack would also work on PJM's feed is an expectation, supported by measurements of PJM's fields and fleet but not by an attack scored against PJM's units.

\begin{enumerate}
\item \textbf{Masking fails.} Matching masked offers against EIA-860, EIA-923 and EPA CEMS puts \R{reid-units-2025} of the \R{reid-n-2025} units we can label in 2025 at the right plant, covering \R{reid-mw-2025} of their capacity, and \R{reid-top5-2025} within the top five guesses. We tune the method on 2024 and test it unchanged on 2025 and 2026 (Section~\ref{sec:reid}). PJM's feed carries every field the method reads, and public data separate PJM's fossil units about as well as ERCOT's, so we expect the result to carry over (Section~\ref{sec:pjm}).
\item \textbf{Exact aggregates leak.} An attacker who knows the other units in a fuel group recovers a target unit's share of high-price offers exactly. An outsider who knows no other unit's offers, only the target's outage dates, still reaches a correlation of \R{out-exact-2025} [\R{out-exact-2025-lo}, \R{out-exact-2025-hi}]. Most of this comes from a few renewable and hydro units that offer most of their capacity at \$250/MWh or more. Over fossil units, whose outages CEMS reveals, the correlation is \R{out-fossil-corr-2025}. A large portfolio owner that knows its own units does no better than the outsider (Section~\ref{sec:aggregates}).
\item \textbf{A private release.} We publish monthly totals by fuel group and price bin with Gaussian noise under zero-concentrated DP, calibrated so that each unit's year is the protected unit. The release has an unbiased noisy layer and derived products computed from it at no further privacy cost. A plant-level variant roughly halves what an attacker learns about a multi-unit plant (Section~\ref{sec:design}).
\item \textbf{A utility test over real uses.} At $\eps=10$ per unit-year the release keeps the market supply curve within \R{curve-twolayer-10}, the monthly clearing price, the supply slope and combined-cycle capacity. A closed-form ratio of a quantity's true spread to the noise predicts the accuracy of every sum-of-cells use to within \R{use-pred-maxgap}. Hourly scarcity and unit-level questions do not survive any monthly release (Section~\ref{sec:eval}).
\end{enumerate}
\looseness=1 Taken together, these results show that masking is not the middle ground between transparency and confidentiality that it appears to be. Public data alone put \R{reid-units-2025} of the masked units we can label at the right plant in 2025 and \R{reid-units-2026} in 2026, and \R{reid-tech-gas-steam-2025} of gas steam units in 2025, so for a substantial share of the fleet a masked release is a named release with a delay. Exact totals protect a unit only from attackers who know little about the other units in them. An operator that wants offers to stay confidential has to add noise, and our release shows what such an operator can still publish.

{}
{}
\section{Background}
\label{sec:background}

\subsection{ERCOT's 60-day disclosure}
\label{sec:ercot}
ERCOT's 60-day day-ahead market disclosure~\cite{ercot60day} holds one record per generation resource per hour. Each record names the resource, its type (for example combined-cycle gas, coal, wind), the qualified scheduling entity (QSE), which is the market participant that submitted it, the upper and lower sustainable limits, start-up and minimum-energy costs, an offer curve of up to ten MW and price points, and the awarded quantity. Combined-cycle plants appear once per configuration. We exclude storage throughout, because ERCOT moved storage resources out of this file in December 2025.

\subsection{PJM's masked disclosure}
\label{sec:pjm-masking}
\looseness=1 PJM publishes the corresponding records in its Energy Market Generation Offers feed~\cite{pjmoffers}, four months after the operating day. Each row covers one unit and one hour. It carries the unit's offer curve of up to 20 MW and price points, its no-load cost and cold, intermediate and hot start costs, its maximum daily starts and minimum run time, and the highest, lowest and average economic maximum and economic minimum of the day. The row does not carry the unit's name, owner, fuel, location or awards. In their place PJM gives each unit an alphanumeric code; the feed's documentation states that it masks generator identification to keep offers confidential and that it changes the codes annually. In the days of the feed we inspected, every listed unit appears in all 24 hours, and a unit with no capacity to offer appears with an economic maximum of zero rather than dropping out of the feed.

The documentation does not say how PJM generates the codes or on which date it changes them. When we say that we mask ERCOT's records the way PJM does, we therefore mean a release with the documented properties, built as follows. We keep only the fields that both operators publish: ERCOT's offer curve, its start-up and minimum-energy costs, and the daily maximum, minimum and mean of its upper and lower sustainable limits, which play the role of PJM's economic maximum and minimum. We replace each resource's name with a keyed hash of the name and the calendar year, so that a unit's code reveals nothing about the unit and codes from different years cannot be linked to each other. Finally, we sort the rows by hour and code, so that ERCOT's alphabetical order of names does not survive. We call the result the \emph{masked release}. Where PJM's documentation leaves a choice open, we make the choice that gives the attacker less: the masked release has fewer offer points and fields than PJM's feed and codes that carry no information, so an attack that succeeds on it uses nothing PJM withholds. We looked at PJM's feed only to establish its format and did not match any PJM unit to a plant. Appendix~\ref{sec:limits} discusses the ethics of the study as a whole.
{}
\subsection{Differential privacy}
\looseness=1 A randomized mechanism $M$ is $(\eps,\delta)$-differentially private if for every pair of neighbouring datasets $D, D'$ and every set of outputs $S$, $\Pr[M(D)\in S] \le e^{\eps}\Pr[M(D')\in S]+\delta$~\cite{dwork2014algorithmic}, where $\eps$ bounds how much one subject's data can change the probability of any output and $\delta$ is a small probability with which that bound may fail. The choice of neighbouring datasets determines what the guarantee protects. We call $D$ and $D'$ neighbours if they differ by adding or removing all of one generating unit's offers for one calendar year, so the guarantee covers everything a unit offers in a year, including its typical price profile, and not only a single day or hour (Section~\ref{sec:privacy-unit} explains this choice).

\looseness=1 We work in zero-concentrated DP ($\rho$-zCDP)~\cite{bun2016concentrated}, whose budgets add up across releases. Because the release repeats every month, this additive composition gives a tighter year-level guarantee for Gaussian noise than composing $(\eps,\delta)$ guarantees directly. The $\ell_2$ sensitivity $\Delta_2$ of a statistic is the largest change in its $\ell_2$ norm when one subject is added or removed, and adding Gaussian noise of standard deviation $\sigma = \Delta_2/\sqrt{2\rho}$ to the statistic gives $\rho$-zCDP. Releasing $n$ statistics with budgets $\rho_i$ gives $\sum_i \rho_i$-zCDP, and $\rho$-zCDP implies $(\rho + 2\sqrt{\rho \ln(1/\delta)},\delta)$-DP. We report every guarantee as $\eps$ at $\delta = \R{delta}$. A guarantee for one unit extends to a group of $k$ units at $\rho k^2$, which matters for plants with several units. Any computation on a released statistic that does not touch the data again keeps the same guarantee, a property known as post-processing~\cite{dwork2014algorithmic,bun2016concentrated}.
{}
\section{Data and threat model}
\label{sec:threat}
This section describes the data we use, how we know which plant each unit belongs to, what a unit's owner may want to keep private, and what the attackers we consider know.

\paragraph{Data.} We use ERCOT's 60-day disclosure for every operating day from January 2024 to September 2026 (\R{use-months} months). Section~\ref{sec:ercot} describes its records. In 2025 an average of \R{units-per-day-2025} resources offered each day, and \R{units-total-2025} distinct resources offered at some point in the year. We sort ERCOT's resource types into five fuel groups: combined cycle, combustion turbine, steam (gas, coal and nuclear), renewables and hydro, and other. These groups are coarse enough that each holds many units and fine enough that the units in a group compete at similar costs. The attacker's side information is all public and national in scope. EIA-860 lists each generator's capacity, minimum load and the dates it entered or left service~\cite{eia860}, EIA-923 reports each plant's monthly net generation~\cite{eia923}, and EPA's Continuous Emissions Monitoring System (CEMS) reports the hourly operation and gross load of every fossil unit above 25~MW~\cite{epacems}. We use 2024 to choose every attack parameter and report 2025 and 2026 as test years, so that no reported result depends on a choice made after seeing it.
{}
\paragraph{Ground truth.} ERCOT names each resource, and many names contain an abbreviation of the plant and the unit number, which lets us link a resource to its EIA plant code. We matched names to plant codes and kept only matches whose name evidence we audited by hand, discarding any resource whose name could fit more than one plant. A wrong label would count a correct guess as an error or a wrong guess as a success, so we prefer fewer labels that are all correct. These labels score the attacks and are never shown to them. They cover \R{reid-n-2025} units in 2025 and \R{reid-n-2026} in 2026.
{}
\paragraph{Secrets.} We consider three things a unit's owner may want kept private. The first is which plant a masked unit belongs to, since that link attaches every offer in the release to a named owner. The second is the days on which a unit was offline, which reveal its maintenance schedule and forced outages. The third is the unit's price profile, which we score as the share of its offered MW priced at \$250/MWh or more. Capacity offered at such prices by a unit that could run more cheaply is the pattern a withholding analysis looks for, so this share is the part of an offer curve that such an analysis examines.
{}
\paragraph{Attackers.} The re-identification attack of Section~\ref{sec:reid} uses public data only. For attacks on a unit's offers we consider four attackers, ordered from the one who knows least to the one who knows most.
\begin{itemize}
\item The \emph{outsider} knows the public data and the days on which the target was offline, which CEMS reveals for any fossil unit because it records the hours in which each unit operated. It knows nothing about any other unit's offers.
\item The \emph{portfolio insider} is a large QSE, or a coalition of the three largest, that knows its own units' offers exactly and can subtract them from every published total.
\item The \emph{partly informed} attacker knows the offers of a fraction $f$ of the other units in the target's fuel group, measured by MW. This attacker stands for an analyst or rival who has estimated most competitors' offers from their costs.
\item The \emph{informed} attacker knows every other unit's offers exactly. This is the worst case DP protects against, and it is the attacker against which we report the private release.
\end{itemize}
We score these attacks by how well the attacker's estimate of each target's high-price share tracks the truth across targets, measured by the correlation between estimated and true shares. Correlation ignores any constant scale or offset in the estimate, so it credits an attacker who ranks units correctly even when its estimates are biased, and ranking units is what a screen for withholding needs. We compute the correlation within each fuel group, average across groups, and give 95\% bootstrap intervals over target units.
{}
\section{Attacks on today's releases}
\label{sec:attacks}
Operators publish offer data today in one of two styles, as unit records with the names masked, as PJM does, or as totals over many units, as ERCOT does in its two-day reports and European exchanges do in their aggregated curves (Section~\ref{sec:intro}). We first attack a masked release that we build from ERCOT's data by PJM's rules (Section~\ref{sec:reid}) and ask whether the result would carry over to PJM's own feed (Section~\ref{sec:pjm}). We then attack exact totals built from the same data (Section~\ref{sec:aggregates}).

\subsection{Re-identifying masked units}
\label{sec:reid}

In the masked release the attacker sees, for each code, a year of hourly offer curves together with the unit's daily maximum and minimum sustainable limits. These limits are close to the capacity and minimum load that EIA-860 lists for the generator, the days with no offered capacity are the unit's outages, and the offered capacity varies by month as the unit's operation does. For each masked code the attacker lists as candidates every EIA-860 plant with a generator whose capacity is within a factor of $e^{0.5}\approx 1.65$ of the code's largest offered capacity. Combined-cycle configurations bundle several turbines, so a candidate can also be a set of up to four generators at one plant. The attacker then compares the code with each candidate through six signatures:
\begin{enumerate}
\item \emph{Static.} The log ratio of the unit's yearly maximum offered capacity to the EIA capacity, the same for minimum load, and whether both lie within a tolerance of 1\%.
\item \emph{Monthly pattern.} Monthly offered capacity against EIA-923 monthly net generation, each divided by its mean and with the fleet-wide seasonal profile removed, then correlated.
\item \emph{Outage days.} The share of the unit's zero-offer days on which a CEMS unit at the candidate plant also did not run, above that CEMS unit's base rate.
\item \emph{Output level.} On days a CEMS unit ran near full load, the log ratio of its peak gross load to the masked unit's economic maximum, which should be near zero and steady for the right plant.
\item \emph{Heat rate.} The unit's offer price near its economic maximum, relative to the fleet's, against the heat rate that CEMS records for the candidate plant's units at high load. A unit's offer price follows its fuel cost, so a unit that burns more fuel per MWh offers higher prices.
\item \emph{Timing.} For a candidate that entered or left service during the year, the gap between the code's first or last offer and the date in EIA-860.
\end{enumerate}

The signatures differ in how much they reveal, and many are missing for some pairs. CEMS covers only fossil units, EIA-860 gives no minimum load for some generators, and the timing signature exists only for plants that opened or closed. Rather than add the signatures with hand-picked weights, the attacker learns how much each one shifts the odds that a candidate is the right plant. A boosted-tree classifier, trained on 2024's pairs of code and candidate with the label whether the candidate is the code's plant, estimates that probability for every pair and treats a missing signature as missing rather than as a mismatch. A plant's score is the probability of its best candidate, the attacker's guess is the plant with the highest score, and its confidence is that plant's share of the summed scores.

A second stage uses the other units at the same plant. One owner usually prepares the offers for all its units at a plant together, so codes at one plant often offer identical prices on the same day, and their outages and economic maxima move together. A second classifier, also trained on 2024, estimates from these features how likely two codes are to share a plant. Each code then also receives the scores that its likely siblings gave each plant, and its capacity is compared with the summed capacity of itself and its siblings, which helps where EIA reports as one generator what ERCOT offers as several units. We choose the model, the tolerance and every other setting on 2024 alone and score 2025 and 2026 once. Their codes are new, so the attacker cannot reuse any match from the training year.

Training needs one year in which the attacker knows which code belongs to which plant. Against ERCOT the named release supplies such a year. Against an operator that masks every year, an attacker could train on ERCOT's named data instead, because every signature is a ratio or a correlation whose meaning does not depend on the market, or on the codes that the static signature alone pins to a single plant. We have not tested either route.

\begin{figure}[tbp]
\centering
\includegraphics{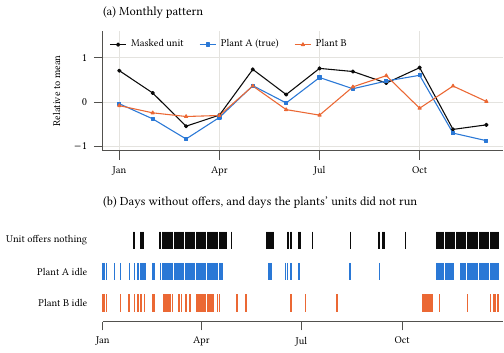}%
\caption{One masked combined-cycle unit of about \R{ex-cap}~MW in 2025, compared with its true plant (A) and with the plant that fits its capacity best (B).}
\label{fig:example}
\end{figure}

Figure~\ref{fig:example} follows one combined-cycle unit of about \R{ex-cap}~MW through the attack in 2025. Its capacity window holds \R{ex-nwin} candidate plants, and at \R{ex-ntol} of them a generator or set of generators matches the unit's capacity and minimum load to within 1\%. The true plant, which we call A, is not among these \R{ex-ntol}. Its capacity differs from what the unit offered by \R{ex-gap-a}, so capacity alone ranks it \R{ex-static-rk}, behind plant B and others that fit almost exactly. The time series separate the two plants. The unit's monthly pattern correlates at \R{ex-r-a} with plant A's monthly generation and at \R{ex-r-b} with plant B's (Figure~\ref{fig:example}a). The unit offered nothing on \R{ex-off-days} days. A CEMS unit at plant A did not run on \R{ex-off-in-a} of them, although it sat idle on only \R{ex-base-a} of all days, while the best-matching unit at plant B did not run on \R{ex-off-in-b} of them, against a base rate of \R{ex-base-b} (Figure~\ref{fig:example}b). Plant A also ranks first on output level and heat rate, and \R{ex-ev-rk} on outage days, where two other plants match slightly better. The classifier ranks plant A first, with \R{ex-post} of the summed score.

The example is typical in one respect. Among the \R{ex-n-fossil} fossil units the attack places correctly in 2025, capacity alone ranks the true plant first for \R{ex-single-static} of them, the monthly pattern for \R{ex-single-r} and the outage days for \R{ex-single-ev}. No single signature names most plants, and the attack works by combining signatures that each point at the right plant only some of the time. Learning how to combine them is what lifts it above a hand-weighted sum. With the first four signatures added under weights tuned on 2024, the top guess is right for \R{reid-v6-2025} of labeled units in 2025. The classifier raises this to \R{reid-v9-2025}, and the sibling stage to \R{reid-units-2025}. The sibling stage's gain is within sampling noise, and in 2026 it does not recur (\R{reid-v9-2026} without the stage, \R{reid-units-2026} with it).

{}

\begin{figure}[tbp]
\centering
\includegraphics{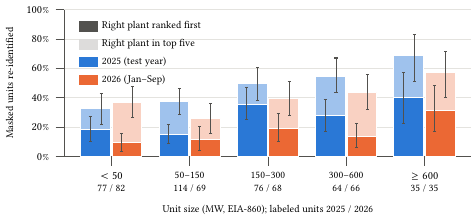}%
\caption{Re-identification of masked units by unit size: share of labeled units whose top guess (dark) or top five guesses (light) include the right plant, 2025 (test year) and 2026. Numbers under each band give labeled units in 2025 and 2026. Whiskers show 95\% bootstrap intervals over units.}
\label{fig:reid}
\end{figure}

\begin{figure}[tbp]
\centering
\includegraphics{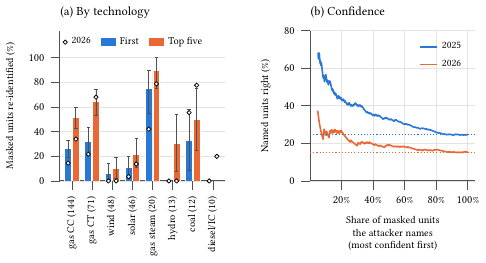}%
\caption{Re-identification by technology (2025 bars, 2026 diamonds), and precision when the attacker names only the units whose best plant beats the runner-up by the widest margin; dotted lines show the precision when it names every unit.}
\label{fig:reidtech}
\end{figure}

Figure~\ref{fig:reid} shows the result by unit size. In 2025 the top guess is right for \R{reid-units-2025} of labeled units, which hold \R{reid-mw-2025} of labeled capacity, and the right plant is among the top five for \R{reid-top5-2025}. In 2026 the rates are \R{reid-units-2026}, \R{reid-mw-2026} and \R{reid-top5-2026}. The lower rates in 2026 come mostly from the public data the attacker has for that year. EIA publishes the full EIA-923 survey only after a year ends, so for 2026 we have only its monthly sample, which covers \R{eia923-tx-plants-2026} Texas plants through August, against \R{eia923-tx-plants-2025} plants over twelve months in the 2025 file. For \R{absent-share-2026} of the units we can label in 2026, the true plant is not in the sample, so the monthly-pattern signature is missing for exactly the right candidate, a case the classifier almost never saw in 2024. This affects \R{absent-share-gas-steam-2026} of gas steam units and \R{absent-share-gas-ct-2026} of combustion turbines, and the attack places only \R{absent-top1-2026} of the affected units correctly. To separate these limits from other differences between the years, we score 2025 again with the models trained on 2024 unchanged and only 2025's data cut to what 2026 offers. Keeping only the plants in the 2026 sample lowers the top-1 rate from \R{cf-base-top1} to \R{cf-sample-top1}. Cutting the offers and CEMS data at 23 September and EIA-923 after August lowers it to \R{cf-window-top1}. Both together lower it to \R{cf-both-top1}, and the top-five rate to \R{cf-both-top5}, below the \R{reid-top5-2026} we measure in 2026. The data limits therefore account for \R{cf-drop-share} of the drop in the top-1 rate and all of the drop in the top-five rate. They reflect when we ran the attack rather than what masking protects, and an attacker who waits for EIA's annual 2026 survey regains the 2025 conditions. Units above 150~MW are matched about twice as often as smaller ones, because few plants have a generator that large, while many plants have small combustion turbines of standard sizes that the static signature cannot tell apart. By technology (Figure~\ref{fig:reidtech}a) gas steam units fare worst for the owner at \R{reid-tech-gas-steam-2025} in 2025, followed by coal (\R{reid-tech-coal-2025}), combustion turbines (\R{reid-tech-gas-ct-2025}) and combined cycle (\R{reid-tech-gas-cc-2025}). Wind and solar stay near zero (\R{reid-tech-wind-2025} and \R{reid-tech-solar-2025}), since CEMS does not cover them and many farms share similar sizes. An attacker need not name a plant for every code. If it names only the matches whose best plant is far ahead of the second best, its accuracy on those matches is well above these averages (Figure~\ref{fig:reidtech}b), and a few confident matches suffice to attach a named owner to a year of offers.

{}

{}
{}
\subsection{Would the attack work on PJM's feed?}
\label{sec:pjm}
We did not run the re-identification attack on PJM's data, and without PJM's unit names there would be no ground truth against which to score it, so we do not report a re-identification rate for PJM. What we can check are the two conditions on which the ERCOT result rests. The first is that PJM's feed carries every field the attack reads. The second is that PJM's generators are about as easy to tell apart from public data as ERCOT's, since a larger or more uniform fleet would leave each masked unit with more plausible plants.

Table~\ref{tab:pjm} pairs each signature with the fields it reads. PJM's feed carries all of them. Its daily economic maximum and minimum supply the static and output-level signatures, its offered MW summed by month supplies the monthly pattern, and a unit that cannot run appears with an economic maximum of zero, which marks the days the outage signature compares with CEMS. Its offer prices supply the heat-rate signature and the identical offers that link units at one plant, and the first and last days on which a code appears supply the timing signature. The side data the attacker matches against, EIA-860, EIA-923 and CEMS, are national datasets and cover PJM's plants exactly as they cover ERCOT's. PJM's four-month delay does not change this, because EIA and EPA publish monthly generation and hourly operation within a few months as well, and the attack only needs both sources for the same past year. PJM's feed also carries fields that the masked release lacks, which can only add to what an attacker learns. Finally, PJM's codes are not random strings. In the days we inspected, a unit's code kept most of its characters from one year to the next, and units that offered identical limits on the same day had codes that differed only in their last characters. The masked release gives the attacker neither of these links, so on this point too it is the harder case.

\begin{table}[tbp]
\caption{The fields each re-identification signature of Section~\ref{sec:reid} reads, in the masked release and in PJM's feed. PJM's feed also carries fields the masked release lacks: 20 offer points instead of ten, no-load cost, maximum daily starts and minimum run time.}
\label{tab:pjm}
\small
\begin{tabular}{@{}>{\raggedright\arraybackslash}p{0.22\columnwidth}>{\raggedright\arraybackslash}p{0.4\columnwidth}>{\raggedright\arraybackslash}p{0.3\columnwidth}@{}}
\toprule
Signature & Masked release (ERCOT) & PJM feed \\
\midrule
Static & daily maximum and minimum of the sustainable limits & Max ECOMAX, Max ECOMIN \\
Monthly pattern & offered MW by month & offered MW by month \\
Outage days & days with no offered MW & days with zero ECOMAX \\
Output level & daily maximum of the upper limit & Max ECOMAX \\
Heat rate & offer price near the upper limit & offer price--MW points \\
Timing & first and last day with offers & first and last day with offers \\
Same-plant links & identical offers on the same day & offer price--MW points \\
Unit code & keyed hash, new each calendar year & changed annually, partly kept across years \\
\bottomrule
\end{tabular}
\end{table}
{}

For the second condition we compute the two strongest signatures on public data alone, separately for the generators in each balancing authority, without any offer data. Among the operable gas combined-cycle, gas combustion-turbine and steam generators in EIA-860 for 2025, \R{fleet-ercot-unique} of ERCOT's \R{fleet-ercot-gens} and \R{fleet-pjm-unique} of PJM's \R{fleet-pjm-gens} have a summer capacity and minimum load that no generator at another plant in the same authority matches to within 5\%, and the median generator shares that range with \R{fleet-ercot-anon} plant in ERCOT and \R{fleet-pjm-anon} plants in PJM, counting its own. CEMS separates units more sharply. Over the first nine months of 2026, \R{fleet-ercot-cems-unique} of ERCOT's \R{fleet-ercot-cems} CEMS units and \R{fleet-pjm-cems-unique} of PJM's \R{fleet-pjm-cems} ran on a pattern of days that no other unit in the same authority shared, and the nearest other unit's pattern differed on a median of \R{fleet-ercot-cems-gap} days in ERCOT and \R{fleet-pjm-cems-gap} in PJM. PJM's larger fleet therefore makes a unit's capacity slightly less distinctive than in ERCOT, while its days of operation remain almost always unique. Both conditions hold, so we expect the attack to identify a similar share of PJM's fossil units. We state this as an expectation supported by these measurements, not as a measured rate.
{}
\subsection{Exact aggregates leak}
\label{sec:aggregates}

\begin{figure}[tbp]
\centering
\includegraphics{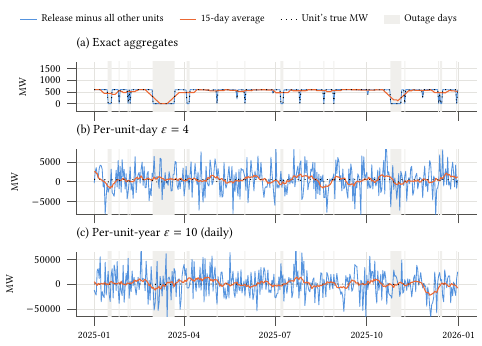}%
\caption{One \R{outage-unit-mw}~MW combined-cycle unit's outages in 2025, seen by the informed attacker through (a) the exact total of its fuel group (Section~\ref{sec:aggregates}) and through the raw layer of two private releases (Section~\ref{sec:rq1}): (b) the per-unit-day release at $\eps=4$ per day and (c) the per-unit-year release at $\eps=10$ published daily.}
\label{fig:outage}
\end{figure}

The totals that operators publish today cover a whole market or a large region~\cite{ercot2day,nordpoolcurves}, which is too coarse for several of the uses of Section~\ref{sec:eval}, since questions about the marginal fuel, the marginal emissions rate or combined-cycle capacity need a total for each fuel. We therefore study the finer release that an operator would publish if it dropped unit records but still wanted to serve those uses, namely exact daily totals for each fuel group, giving the MW offered in each of seven price bins and the share of offers in each bin averaged over units. A total is a sum, however, and anyone who knows all terms but one can recover the remaining term. An informed attacker subtracts every other unit's contribution and is left with the target's own vector, so it recovers the target's high-price share with correlation \R{tab-unit-exact} and its outage days exactly (Figure~\ref{fig:outage}a). Group size does not prevent this. The combined-cycle and combustion-turbine totals would satisfy the 15/15 rule on \R{rule1515-cc-days} and \R{rule1515-ct-days} of days in 2025, with a median of \R{rule1515-cc-n} and \R{rule1515-ct-n} units, and they leak as completely as the smaller groups.

{}

An attacker who knows much less can still learn a great deal. The outsider compares the group's share vector on the target's outage days with the same vector on the days just before and after each outage. The other units change little from one day to the next, so whatever disappears from the total when the target goes offline is mostly the target's own profile. This reaches a correlation of \R{out-exact-2025} [\R{out-exact-2025-lo}, \R{out-exact-2025-hi}] in 2025 and \R{out-exact-2026} in 2026. Both figures pool the fuel groups, and most of what they measure comes from one group. Among renewables and hydro, a few units offer at least half their capacity at \$250/MWh or more while the rest offer almost nothing there, and the outsider tells the two kinds apart well, with a correlation of \R{out-ren-corr-2025} within the group in 2025 and \R{out-ren-corr-2026} in 2026. Combined-cycle units and combustion turbines offer almost nothing at those prices in either year, so their groups add estimation noise but little true spread, and over the combined-cycle, combustion-turbine and steam groups the correlation is \R{out-fossil-corr-2025} in 2025 and \R{out-fossil-corr-2026} in 2026. CEMS records the outages of these fossil units but not of renewable or hydro units, so an outsider who learns outage dates from CEMS alone reaches the lower figure, and the higher one requires outage dates from other sources, such as outage notices. This accounts for most of the drop between the years. The group holds \R{out-ren-hi-units-2025} targets that offer at least half their capacity at high prices in 2025, which supply \R{out-ren-cov-share-2025} of the pooled covariance, and only \R{out-ren-hi-units-2026} in 2026, so the group in which the outsider does well carries less weight. A market-wide total of the kind published today contains many more units than a fuel group, and their day-to-day changes would add more noise to this contrast, so the outsider would do worse against it. We have not measured how much worse.

\begin{figure}[tbp]
\centering
\includegraphics{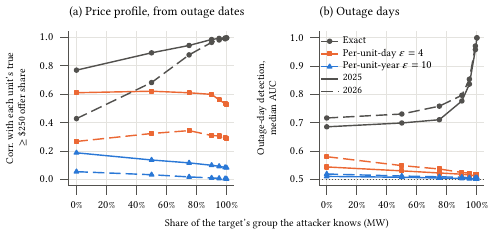}%
\caption{Attack success against the share $f$ of the target's fuel group the attacker knows on exact daily totals (Section~\ref{sec:aggregates}) and on the raw layer of two private releases (Section~\ref{sec:rq1}). The per-unit-year release is published daily.}
\label{fig:knowledge}
\end{figure}

Figure~\ref{fig:knowledge} sweeps the share $f$ of the group the attacker knows. Correlation climbs from \R{ks0-exact-2025} at $f=0$ to \R{ks75-exact-2025} at $f=0.75$ and \R{ks90-exact-2025} at $f=0.9$. An attacker that does not already know the outage days has a harder task in finding them from the totals. The area under the ROC curve for detecting them is \R{ks-exact-auc-f0} at $f=0$ and reaches \R{ks-exact-auc-f99} only at $f=0.99$.

{}

A third attacker sits between these two. A large portfolio owner, which in ERCOT submits offers as a QSE, knows its own units' offers exactly and can subtract them from every published total before it looks for a target among the rest. One might expect this to help a great deal, since the three largest QSEs together offer a median \R{qse3-known-2025} of the other MW in a target's group, and Figure~\ref{fig:knowledge} shows that an attacker who knows that share of a group's MW in general does much better than the outsider. We tested this by giving the coalition of the three largest QSEs the outsider's attack on the totals minus their own units, with targets drawn from the units they do not schedule. The coalition does no better than the outsider. Its correlation exceeds the outsider's on the same targets by \R{qse3-gain-2025} [\R{qse3-gain-2025-lo}, \R{qse3-gain-2025-hi}] in 2025, and in 2026 it is lower by a small margin (a difference of \R{qse3-gain-2026}). The reason lies in how the outsider's estimate is formed. It contrasts the total on the days around each outage with the total on the outage days, so a unit whose offers are the same on both kinds of day cancels out of the estimate without being subtracted. Removing the coalition's units therefore changes the estimate only by the amount their own offers change between those days. In the bins at or above \$250/MWh, where the target's high-price share is measured, that change is \R{qse3-hicontrast-2025} of the group's contrast for the median target in 2025 (\R{qse3-hicontrast-2026} in 2026). What helps an attacker is knowing the units whose offers move around the target's outages, and the large portfolios' units are not those units.

The two release styles in use therefore fail in different ways. Masking leaves identities exposed, because a unit's limits and operating days match public records. Aggregation hides identities but leaves prices exposed, because anyone who knows or can model the other terms of an exact total can difference it, and even the outsider's crude model of the other units suffices. No choice of grouping removes this weakness, so a release that protects offers must add noise whose scale does not depend on what the attacker knows.
{}
\section{A private release}
\label{sec:design}
Section~\ref{sec:attacks} showed that neither masking nor exact totals protect a unit's offers, and that any exact total can be differenced by an attacker who knows enough about the other units. A release that protects offers against every attacker must therefore add noise, and the question is how to add it so that the release stays useful. The release is meant for operators that publish offers but want them to stay confidential, as PJM does today, and not for ERCOT, which has chosen to publish names. We use ERCOT's data only because its names let us measure what the release protects. This section describes what we release (Section~\ref{sec:released}), which unit of data the guarantee protects and how often we publish (Section~\ref{sec:privacy-unit}), the noise mechanism (Section~\ref{sec:mechanism}), how we handle the negative values that noise produces (Section~\ref{sec:two-layer}), and the formal accounting (Section~\ref{sec:accounting}).

\subsection{What is released}
\label{sec:released}
The release replaces unit records with totals by month, fuel group and price bin, and it publishes two such totals, one measuring capacity and one measuring units' pricing behaviour. The price bins are fixed and public: below \$0, \$0 to 25, 25 to 50, 50 to 100, 100 to 250, 250 to 1{,}000, and \$1{,}000/MWh or more. We fix them in advance because bins chosen from the data would themselves reveal something about the offers and would have to be paid for from the privacy budget.
\begin{itemize}
\item \emph{Offered MW.} For each hour, the MW each unit offers in each bin, summed over the group's units and averaged over the month's hours. Each unit-hour's offer is capped at 600~MW before summing, so no unit contributes more than 600~MW to any hour. The cap sets the scale of the noise, which grows in proportion to it, while only units larger than the cap lose any offered capacity. A cap near the size of the largest single units therefore trades a small loss at the top of the supply curve against noise that would otherwise be set by the very largest unit. This statistic gives the market supply curve and the capacity offered at each price.
\item \emph{Unit share.} For each hour, each unit adds the fraction of its own offer that falls in each bin, a vector that sums to one. These are summed over units and averaged over the month. This statistic counts units instead of MW, so small units are not drowned out by large ones, and its high-price cells show how many units priced part of their capacity high, which is the quantity a withholding screen examines.
\end{itemize}
Five groups and seven bins give 35 cells per statistic per month. Removing one unit changes a month's MW cells by at most 600~MW in $\ell_2$ norm and its share cells by at most 1, because an hour's offer splits across bins and averaging over hours cannot increase the norm (Section~\ref{sec:accounting}).
{}
\subsection{Privacy unit and cadence}
\label{sec:privacy-unit}
The natural first design protects one unit-day and publishes daily, matching the cadence of the current disclosure. The resulting guarantee is weak, because a unit offers on most days of the year and the privacy losses of the daily releases add up over the days on which it appears. A per-day $\eps=4$ is $\eps=\R{day-year-eps-4}$ for the unit's year, and per-day $\eps=1$ is still $\eps=\R{day-year-eps-1}$. What a competitor wants is a unit's typical price profile, which every daily release reveals a little more of, so the guarantee that matters is the year-level number.

We therefore protect each unit's calendar year and publish monthly. The year's budget is split evenly across the 12 monthly releases and between the two statistics. Averaging over a month before adding noise is what makes this affordable. Each monthly release receives a twelfth of the year's budget instead of a 365th, and under zCDP the noise scale grows with the square root of the number of releases, so a daily release under the same yearly guarantee would need $\sqrt{365/12}\approx 5.5$ times more noise per cell.

Differential privacy does not say which value of $\eps$ is safe. The budget bounds how much one unit's year of offers can change the probability of any output, by a factor of at most $e^{\eps}$. At the budgets used in practice this worst-case bound is loose, since real attacks learn far less than it allows, so deployed releases choose a budget by weighing what the data must still support against what attacks still learn. The 2020 US Census, for example, released its redistricting data at $\eps=17.14$ for persons~\cite{censusparams2021}. That budget covers a person's records in a release made once a decade, while ours covers one year of a unit's offers and renews every year, so the two are not directly comparable, but they show the range in which deployed releases operate. We therefore report every design over a grid of budgets from $\eps=1$ to $\eps=30$ per unit-year and choose the two budgets we discuss in detail, $\eps=4$ and $\eps=10$, from what the grid shows about the uses of the release and about the attacker (Section~\ref{sec:budget}).

A plant with several units is a group of units, and a per-unit guarantee weakens with group size: $\eps=10$ per unit is $\eps=\R{group-eps-k2}$ for a two-unit plant, \R{group-eps-k3} for three units and \R{group-eps-k4} for four. The \emph{plant-year} variant protects whole plants instead. It scales each plant's contribution down to at most 1{,}200~MW and two units' worth of share per hour, which doubles the sensitivity and so the noise.
{}
\subsection{Mechanism}
\label{sec:mechanism}
We add independent Gaussian noise to every cell. A yearly budget $\eps$ at $\delta=\R{delta}$ corresponds to $\rho=\R{rho-10}$ for $\eps=10$ and $\rho=\R{rho-4}$ for $\eps=4$. Each statistic in each month receives $\rho/24$, so the MW cells get $\sigma = 600/\sqrt{2\rho/24}$: \R{sigma-month-10}~GW per cell at $\eps=10$ and \R{sigma-month-4}~GW at $\eps=4$. For scale, the combined-cycle group offers \R{cc-fleet-gw}~GW in an average month of 2025.

The Laplace mechanism would give pure $\eps$-DP ($\delta=0$). With basic composition over the same releases, it adds \R{laplace-ratio-10} more noise than the Gaussian at $\eps=10$ and \R{laplace-ratio-4} more at $\eps=4$. The difference comes from composition over the 24 releases of a year. Under zCDP the Gaussian mechanism's year-level $\eps$ grows roughly with the square root of the number of releases, whereas under basic composition the Laplace mechanism's grows linearly. We therefore use the Gaussian.
{}
\subsection{A two-layer release}
\label{sec:two-layer}
Noisy cells can be negative, and an offer curve cannot. The obvious fix, clipping each cell at zero, adds an upward bias that grows with the noise~\cite{mcglinchey2020nonnegative,zhu2021bias}. Clipping replaces every negative draw with zero but leaves positive draws unchanged, so a cell whose true value is small is pushed upward on average, and a sum over many such cells, such as a point on the supply curve, accumulates the bias. At $\eps=4$ the clipped supply curve is off by \R{curve-clip-4}, and at $\eps=1$ by \R{curve-clip-1}.

We publish two layers instead, following the noisy measurement files of the 2020 US Census~\cite{abowd2022census}. The \emph{raw layer} holds the noisy cells as drawn, negatives included. It is unbiased, and the error of any sum of cells is Gaussian with a known variance, so a user can attach exact error bars to any linear quantity. The \emph{derived layer} is computed from the raw layer alone and so costs no privacy:
\begin{itemize}
\item a monotone supply curve, by isotonic regression of the cumulative MW across price bins, projected to be non-negative~\cite{lee2015maximum};
\item a clearing price, where the clipped cumulative curve meets the month's average day-ahead MW awarded to resources. That total is already public: ERCOT posts the energy sold at each settlement point after every day-ahead run~\cite{ercotdamsold}, and subtracting the energy-only (virtual) offer awards, which the 60-day disclosure lists in a report of their own~\cite{ercot60day}, leaves the resource total;
\item unit shares projected onto the probability simplex.
\end{itemize}
With the derived curve, the error at $\eps=4$ falls from \R{curve-clip-4} to \R{curve-twolayer-4}, and at $\eps=1$ from \R{curve-clip-1} to \R{curve-twolayer-1}. One alternative is to condition the noise inside the mechanism on a nonnegative output, so that no negative values appear. That costs privacy, turning $\eps=10$ into \R{trunc-eps-10} and $\eps=4$ into \R{trunc-eps-4}, and the output is still biased. Section~\ref{sec:accounting} gives the sensitivity and accounting statements.
{}
\subsection{Privacy accounting}
\label{sec:accounting}

Let $x_{u,t} \in \mathbb{R}_{\ge 0}^{B}$ be unit $u$'s offered MW in each of the $B=7$ price bins in hour $t$, after capping so that $\|x_{u,t}\|_1 \le C = 600$~MW. Let $s_{u,t}\in\Delta^{B}$ be its share vector, zero in hours it does not offer. For month $m$ with hours $T_m$ and fuel group $g$, the released statistics are
\[
  X_{m,g} = \frac{1}{|T_m|}\sum_{t\in T_m}\sum_{u\in g} x_{u,t},
  \qquad
  S_{m,g} = \frac{1}{|T_m|}\sum_{t\in T_m}\sum_{u\in g} s_{u,t}.
\]

\begin{lemma}\label{lem:sensitivity}
Adding or removing one unit's offers for a year changes each month's $X_m$ (all groups stacked) by at most $C$ in $\ell_2$ norm and $S_m$ by at most $1$.
\end{lemma}

\begin{theorem}\label{thm:privacy}
Releasing $X_m + \mathcal{N}(0,\sigma_X^2 I)$ and $S_m + \mathcal{N}(0,\sigma_S^2 I)$ for the 12 months of a year, with $\sigma_X = C/\sqrt{2\rho/24}$ and $\sigma_S = 1/\sqrt{2\rho/24}$, satisfies $\rho$-zCDP with respect to one unit's year, and hence $(\rho+2\sqrt{\rho\ln(1/\delta)},\delta)$-DP. Every derived product in Section~\ref{sec:design} satisfies the same guarantee.
\end{theorem}
Appendix~\ref{app:proofs} proves both.

\paragraph{Plant-year release.} Scaling each plant's hourly contribution to at most $K$ units and $KC$~MW makes the sensitivities $KC$ and $K$; the same theorem holds with a plant's year as the protected unit.

\paragraph{Per-day releases.} A daily release at $\eps_d$ per unit-day, repeated over $n$ days, composes in zCDP to $n\rho(\eps_d)$ for the unit's year; for 2025 this gives the year-level values in Table~\ref{tab:designs}.

\paragraph{Conditioned noise.} The alternative in Section~\ref{sec:two-layer} draws each release from the Gaussian conditioned on every cell being nonnegative. The output density's normalizing constant then depends on the data, and the guarantee weakens.
\begin{proposition}\label{prop:trunc}
With every release of Theorem~\ref{thm:privacy} conditioned on a nonnegative output, the year of releases satisfies $(\xi,\rho)$-zCDP~\cite{bun2016concentrated} with $\xi = 24\sqrt{2/\pi}\,C/\sigma_X = \sqrt{96\rho/\pi}$, and hence $(\xi+\rho+2\sqrt{\rho\ln(1/\delta)},\delta)$-DP.
\end{proposition}
Each release adds \R{trunc-cost-10} to $\eps=10$, so the year-level guarantee becomes \R{trunc-eps-10}; $\eps=4$ becomes \R{trunc-eps-4}. Appendix~\ref{app:proofs} gives the proof.
{}
\section{Evaluation}
\label{sec:eval}

We ask four questions. What does an attacker learn from the private release (RQ1)? How much of the market picture does it keep (RQ2)? Which uses of the data survive, and can we predict which ones in advance (RQ3)? What can no monthly release do (RQ4)? Table~\ref{tab:designs} collects the main designs. Unless stated otherwise, results are for 2025, the private release is the per-unit-year monthly one, and attacks run on the raw layer, which is what the release publishes. Every derived product is a function of the raw layer, so attacking the raw layer covers them too.

\begin{table*}[tbp]
\caption{Release designs, 2025. Noise is the standard deviation per MW cell. Curve error is the relative error of the derived market supply curve, price error the mean absolute error of the implied monthly clearing price. The attack columns give the informed attacker's correlation between estimated and true high-price share, on the raw layer, for one unit and for a whole plant.}
\label{tab:designs}
\small
\begin{tabular}{lrrrrrrr}
\toprule
 & \multicolumn{2}{c}{$\varepsilon$ ($\delta=10^{-6}$)} & Noise & \multicolumn{2}{c}{Release keeps} & \multicolumn{2}{c}{Attacker learns} \\
\cmidrule(lr){2-3}\cmidrule(lr){5-6}\cmidrule(lr){7-8}
Release & per release & per unit-year & $\sigma$/cell (MW) & curve err. & price err. (\$) & one unit & one plant \\
\midrule
Exact aggregates (600 MW cap, no noise) & --- & $\infty$ & 0 & 7\% & 1.8 & 1.00 & 1.00 \\
Per unit-day, daily & 4 & 164 & 1,191 & 9\% & 2.3 & 0.92 & --- \\
Per unit-day, daily & 1 & 25 & 4,540 & 23\% & 10.4 & 0.62 & --- \\
Per unit-year, daily & 10 & 10 & 9,855 & 49\% & 22.8 & 0.33 & --- \\
Per unit-year, monthly & 10 & 10 & 1,787 & 11\% & 3.0 & 0.23 & 0.56 \\
Per unit-year, monthly & 4 & 4 & 4,125 & 25\% & 7.4 & 0.08 & 0.29 \\
Per plant-year, monthly (1,200 MW / 2 units) & 10 & 10 & 3,574 & 20\% & 7.1 & --- & 0.30 \\
\bottomrule
\end{tabular}
{}
\end{table*}
{}
{}
\subsection{Choosing a budget}
\label{sec:budget}

\begin{figure*}[tbp]
\centering
\includegraphics{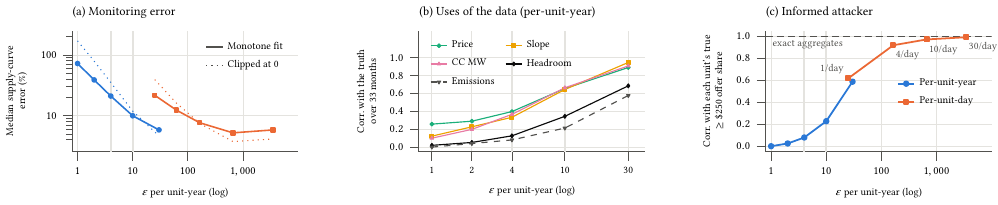}%
\caption{The per-unit-year release as its budget grows, with $\eps$ per unit-year at $\delta=\R{delta}$ on a log axis. Light vertical lines mark $\eps=4$ and $\eps=10$. Per-unit-day releases sit at their year-level $\eps$ and are labelled with the per-day budget. Panels (a) and (c) are for 2025, and (b) covers the \R{use-months} months of 2024 to 2026.}
\label{fig:tradeoff}
\end{figure*}

Figure~\ref{fig:tradeoff} shows how the per-unit-year release behaves as its budget grows from $\eps=1$ to $\eps=30$. Panel (a) gives the error of the derived supply curve, panel (b) how closely five uses of the data from RQ3 follow their true values from month to month, and panel (c) how closely the informed attacker of Section~\ref{sec:aggregates}, who knows every other unit in the target's fuel group, recovers the target's share of offers at or above \$250/MWh. A larger budget means less noise, so every curve in the figure improves as $\eps$ grows, the attacker's as well as those of the uses. What changes along the grid is which of the two gains more from each increase.

\looseness=-1 At small budgets the release protects the units well but supports little. At $\eps=2$ the attacker's correlation is \R{sweep-att-2}, close to zero, but the supply curve is off by \R{curve-twolayer-2}, and the clearing price, the use that holds up best, follows the truth with a correlation of only \R{sweep-use-price-2}. Up to $\eps=10$ each increase in the budget helps the uses more than it helps the attacker. From $\eps=4$ to $\eps=10$ the correlation of the clearing price rises from \R{use-price-4} to \R{use-price-10}, that of the supply slope from \R{use-slope-4} to \R{use-slope-10} and that of the combined-cycle MW offered from \R{use-offered0-4} to \R{use-offered0-10}, while the attacker's rises from \R{tab-month4-unit} to \R{inf-month10-2025}. Beyond $\eps=10$ the balance reverses. From $\eps=10$ to $\eps=30$ the same three uses rise to between \R{sweep-use-price-30} and \R{sweep-use-slope-30}, but the attacker's correlation more than doubles, to \R{sweep-att-30}, which is close to the \R{tab-day1-unit} it reaches against the per-unit-day release at $\eps=1$ per day. The supply curve gains little over this range. Its error falls from \R{curve-twolayer-10} to \R{curve-twolayer-30}, which is close to the \R{curve-cap-only} error that the 600~MW cap causes without any noise, so at $\eps=10$ most of the error that noise can remove is already gone. Headroom below \$250/MWh and the marginal emissions rate track the truth less closely than the other uses at every budget, for reasons that RQ3 explains, and they need budgets above $\eps=10$ to follow it closely.

\looseness=-1 We therefore discuss two budgets in detail and use them as stand-ins for a lenient and a strict operator. The budget $\eps=10$ is the point on the grid where the balance turns, since below it each increase helps the uses more than the attacker and above it each increase helps the attacker more. At this budget the clearing price, the supply slope and the combined-cycle MW offered follow the truth with a correlation of about \R{use-price-10}, and the attacker reaches \R{inf-month10-2025}, against \R{tab-unit-exact} on exact totals. It is our main setting. The budget $\eps=4$ shows what a stricter operator gains and what it gives up. It holds the attacker to \R{tab-month4-unit}, close to zero, but the same three uses fall to between \R{use-slope-4} and \R{use-price-4}, and the supply curve is off by \R{curve-twolayer-4}. We also use $\eps=4$ as the per-day budget of the per-unit-day design, to show that a budget which sounds strict for one day still protects little over a year (RQ1). Neither value is a recommendation (Appendix~\ref{sec:limits}). Where an operator should stop depends on which uses it must serve and on how much it is willing to let an attacker learn, and Figure~\ref{fig:tradeoff} lets it read off both.

{}
{}
\subsection{RQ1: what does an attacker learn?}
\label{sec:rq1}

Figure~\ref{fig:tradeoff}c and the last two columns of Table~\ref{tab:designs} give the informed attacker's correlation. Per-day guarantees protect little: at per-day $\eps=4$ the attacker still reaches \R{inf-day4-2025} [\R{inf-day4-2025-lo}, \R{inf-day4-2025-hi}], and at per-day $\eps=1$ \R{tab-day1-unit}, because the noise in each daily release is drawn independently, so an attacker who averages a year of daily estimates divides the noise's standard deviation by about the square root of the number of days, while the target's price profile, which changes slowly, survives the averaging. The per-unit-year guarantee holds the attacker to \R{tab-yeardaily-unit} with daily publication and \R{inf-month10-2025} [\R{inf-month10-2025-lo}, \R{inf-month10-2025-hi}] with monthly publication at $\eps=10$ (\R{inf-month10-2026} in 2026), and to \R{tab-month4-unit} at $\eps=4$. The outsider does worse still, reaching \R{out-yeardaily10-2025} [\R{out-yeardaily10-2025-lo}, \R{out-yeardaily10-2025-hi}] on the per-unit-year daily release at $\eps=10$ (Figure~\ref{fig:knowledge}a). The unit whose outages the informed attacker recovered in full from exact totals (Figure~\ref{fig:outage}a) shows no visible outages in either private release (Figure~\ref{fig:outage}b, c).

A plant is a different target. The units of one plant usually share an owner and often offer alike, so an attacker who wants to know how a plant prices can treat all of its units in a fuel group as one target, and the per-unit guarantee then covers only part of what the attacker is after. Aiming the same attack at all of a plant's units in a group, the per-unit release at $\eps=10$ allows a correlation of \R{plantatk-unit10-2025} [\R{plantatk-unit10-2025-lo}, \R{plantatk-unit10-2025-hi}] across \R{plantatk-n-2025} plant-group targets, which is the group-privacy loss of Section~\ref{sec:privacy-unit} showing up in practice. The plant-year release at the same $\eps$ cuts this to \R{plantatk-plant2-10-2025} [\R{plantatk-plant2-10-2025-lo}, \R{plantatk-plant2-10-2025-hi}], about what the per-unit release gives at $\eps=4$ (\R{plantatk-unit4-2025}).

A correlation is not a privacy guarantee. The attacks we run are strong but not optimal, and the DP guarantee bounds every attack, including ones we did not try. The measured numbers say how much a realistic attacker gets in practice, which is what a market operator weighing a release design needs to know alongside $\eps$.
{}
\subsection{RQ2: what does the release keep?}
\label{sec:rq2}

\begin{figure}[tbp]
\centering
\includegraphics{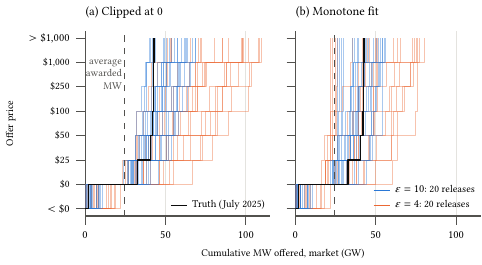}%
\caption{The July 2025 market supply curve against 20 draws of the per-unit-year monthly release. Clipping shifts the curve right; the monotone fit of the same draws stays centred on the truth.}
\label{fig:curve}
\end{figure}

The first thing a reader builds from offer data is the market supply curve. Capping each unit-hour at 600~MW alone, with no noise, already puts the curve \R{curve-cap-only} off, because the largest units offer more than the cap. Adding noise at $\eps=10$ raises the error to \R{curve-twolayer-10}, and at $\eps=4$ to \R{curve-twolayer-4} (Figure~\ref{fig:tradeoff}a, Figure~\ref{fig:curve}). The per-unit-year daily release at the same $\eps=10$ is off by \R{tab-yeardaily-curve}, which is the price of publishing 365 times a year instead of 12.

{}

\begin{figure}[tbp]
\centering
\includegraphics{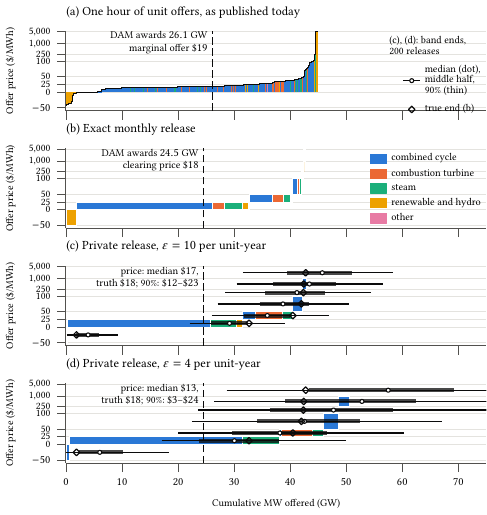}%
\caption{The merit order before and after privacy, July 2025. (a) One hour of unit offers as published today (15 July, hour ending 17), each unit at its own price. (b) The exact monthly release: MW per fuel group and price band, against average DAM awards. (c, d) One private release (blocks) and, over 200 releases, where each price band ends.}
\label{fig:merit}
\end{figure}

\begin{figure}[tbp]
\centering
\includegraphics{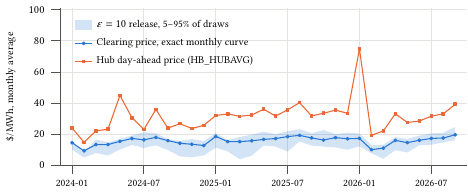}%
\caption{Monthly clearing price implied by the released supply curve against the hub day-ahead price, 2024 to 2026: exact aggregates and the $\eps=10$ release (5 to 95\% band over draws).}
\label{fig:price}
\end{figure}

\looseness=1 Figure~\ref{fig:merit} follows one month through the pipeline. In July 2025 the exact monthly curve clears at \$\R{mo-price-true}/MWh. Across 200 private releases the median clearing price is \$\R{mo-price-10} (90\% of releases between \$\R{mo-price-10-lo} and \$\R{mo-price-10-hi}) at $\eps=10$ and \$\R{mo-price-4} (\$\R{mo-price-4-lo} to \$\R{mo-price-4-hi}) at $\eps=4$. Over all months the mean absolute price error is \$\R{price-twolayer-10}/MWh at $\eps=10$ and \$\R{price-twolayer-4} at $\eps=4$. A clearing price built from monthly offers is only a rough guide to the hub price, even without noise, because averaging a month of offers and demand smooths away the hours of scarcity that drive the average price, and because the day-ahead price also depends on imports, virtual bids and transmission limits that the offer curve does not capture. Their correlation across months is \R{hub-corr-exact} for the exact release and \R{hub-corr-10} at $\eps=10$ (Figure~\ref{fig:price}).

{}
{}
{}
\subsection{RQ3: which uses survive?}
\label{sec:rq3}

\begin{figure}[tbp]
\centering
\includegraphics{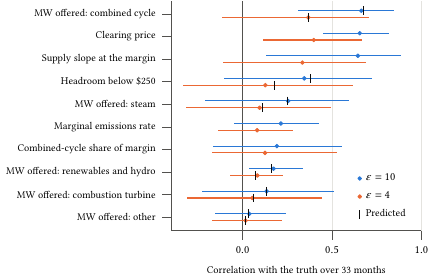}%
\caption{Which uses the release supports. Dots give the median correlation with the truth over 200 releases and bars span 90\% of them. Ticks give the prediction for sums of cells.}
\label{fig:scorecard}
\end{figure}

\begin{figure*}[tbp]
\centering
\includegraphics{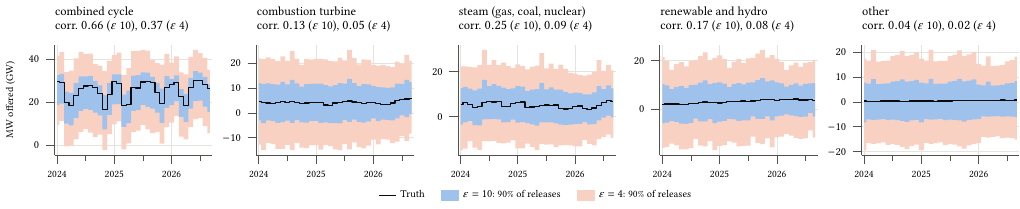}%
\caption{MW offered per fuel group per month, 2024 to 2026: truth against the middle 90\% of 200 private releases at $\eps=10$ and $\eps=4$. The combined-cycle fleet's seasonal pattern survives; the small groups do not.}
\label{fig:offered}
\end{figure*}

RQ2 asked about the supply curve as a whole. Analysts, however, use offer data to compute specific quantities, and a release can keep some of them and lose others. We computed each of the quantities below for every month, once from the truth, the unclipped monthly statistic, and once from each of 200 private releases, and scored each quantity by its correlation with the truth across the \R{use-months} months (Figure~\ref{fig:scorecard}). Quantities that are sums of cells use the raw layer, which is unbiased, while band shares and the slope use the monotone fit for each fuel group, since they need a curve that never decreases.
{}
\begin{itemize}
\item \emph{Offered MW by fuel group}: total MW offered per group (sum of 7 cells).
\item \emph{Headroom}: MW offered below \$250/MWh across groups (25 cells) minus average awarded MW.
\item \emph{Clearing price}: where average demand crosses the clipped market supply curve.
\item \emph{Slope}: MW offered per \$1/MWh in the price band where demand crosses.
\item \emph{Marginal fuel and emissions rate}: each group's share of the crossing band, and the share-weighted emission factor. Factors, in tons of CO$_2$ per MWh, come from typical heat rates times EIA fuel CO$_2$ coefficients: combined cycle 0.40, combustion turbine 0.58, steam 0.59 (weighted by 2025 offered MW of coal, gas steam and nuclear), renewables and hydro 0, other 0.73.
\end{itemize}
{}
\paragraph{A prediction.} Many uses are sums of cells. MW offered by a fuel group sums its 7 bins, and headroom below \$250/MWh sums 25 cells across groups. For such a use, the raw layer adds independent noise of variance $k\sigma^2$ to a quantity whose true values vary across months with standard deviation $\mathrm{sd}$. The expected correlation between truth and release is therefore $r \approx \mathrm{sd}/\sqrt{\mathrm{sd}^2 + k\sigma^2}$. The prediction matches the measured correlation in all 12 linear cases (six uses, two values of $\eps$) to within \R{use-pred-maxgap}. Combined-cycle MW, with a large seasonal swing, reaches \R{use-offered0-10} at $\eps=10$ (predicted \R{use-offered0-10-pred}). Headroom reaches \R{use-headroom-10} (predicted \R{use-headroom-10-pred}). The small groups fail as predicted: combustion turbines at \R{use-offered1-10}, renewables and hydro at \R{use-offered3-10} and the other group at \R{use-offered4-10} (Figure~\ref{fig:offered}). The ratio gives operators and users a way to judge a use before any release: what matters is how much a quantity moves from month to month compared with $\sqrt{k}\,\sigma$. The true spread must come from outside the release, for example from earlier exact data or the published fleet capacity, because the spread of the released values includes the noise itself and estimating the true spread from a single 33-month release is too noisy to be useful.

{}
\begin{figure*}[tbp]
\centering
\includegraphics{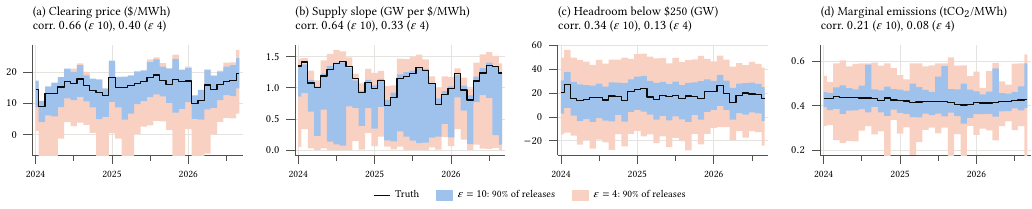}%
\caption{Monthly quantities at the margin: truth against 200 private releases. Headroom is MW offered below \$250/MWh minus demand.}
\label{fig:margin}
\end{figure*}
{}

\paragraph{Non-linear uses.} The clearing price reaches \R{use-price-10} at $\eps=10$ and \R{use-price-4} at $\eps=4$; the supply slope at the clearing point \R{use-slope-10} and \R{use-slope-4} (Figure~\ref{fig:margin}a and b). The marginal emissions rate, which depends on which fuel sits at the margin and how steep the curve is there, reaches only \R{use-mef-10}. The marginal fuel group itself is combined cycle in every month, and the release agrees in \R{use-fuel-agree-10} of month-draws at $\eps=10$, which says more about ERCOT's fleet than about the release.
{}
\subsection{RQ4: what can't a monthly release do?}
\label{sec:rq4}

\begin{figure}[tbp]
\centering
\includegraphics{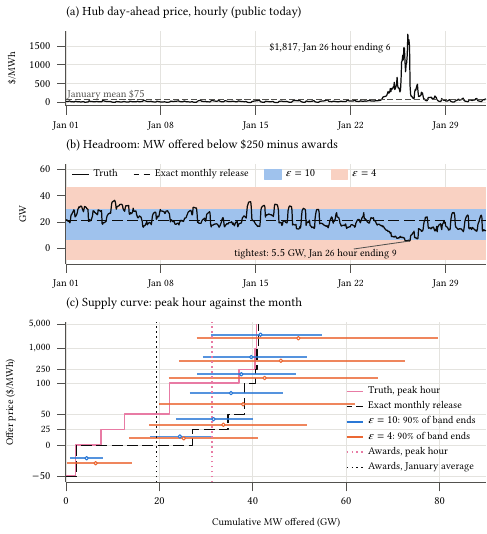}%
\caption{The January 2026 cold snap: the hourly truth against the single monthly number the release gives. The bands in (b) span 90\% of 200 private releases.}
\label{fig:spike}
\end{figure}

Some questions are about hours, not months. During the January 2026 cold snap the hub price peaked at \$\R{spike-hub-peak}/MWh in hour ending 6 on January 26, against a monthly mean of \$\R{spike-hub-mean} (Figure~\ref{fig:spike}a). Hourly headroom below \$250/MWh ranged from \R{spike-head-min} to \R{spike-head-max}~GW over the month, while the monthly release reports a single \R{spike-head-month}~GW. At the peak hour only \R{spike-below50-peak}~GW was offered below \$50/MWh, against \R{spike-below50-month}~GW in the monthly curve. The peak-hour curve clears near \$\R{spike-price-peakcurve}/MWh at that hour's demand, while the monthly curve says \$\R{spike-price-monthcurve}.

{}

This is a loss of resolution, not of privacy budget. The exact monthly release has the same blind spot, and no amount of $\eps$ fixes it. Hourly scarcity studies need an hourly release with its own budget, and studies of single units, such as a withholding case against one firm, need the unmasked data. The market monitor already has that data and does not need a public release for it.
{}
\section{Related work}
\label{sec:related}

\paragraph{Privacy in energy data.} Most privacy work in energy protects households: hiding load with batteries~\cite{laforet2016individual}, DP for load disaggregation~\cite{wang2020privacy}, private clustering of meter data~\cite{ravi2022differentially}, and the cost of fine-grained consumption data~\cite{voyez2025privacy}. DP has also been applied to grid models~\cite{fioretto2020differential}, optimal power flow~\cite{dvorkin2021differentially} and control of partitioned resources~\cite{laufer2024privacy}. A proposed privacy standard for inverter-based and distributed resources argues that grid data rules should state what they protect~\cite{currie2023data}. None of this work treats market disclosures, where the data subjects are a few hundred generators with public plant records.

\paragraph{Market transparency.} Operators publish offer data with different delays, aggregation and masking~\cite{yang2018review}. Offer data underpins studies of market power and strategic bidding~\cite{borenstein2002measuring,hortacsu2008understanding} and recent work on withholding and mitigation~\cite{bansal2022market,zhou2024withholding}. On Alberta data, Brown et al.\ find that aggregate market information predicts firms' bids and firm-specific information adds little~\cite{brown2025disclosure}, which suggests aggregate releases like ours keep what market participants use. We measure what disclosure reveals, not how it changes bids.

\paragraph{Re-identification and post-processing.} Exact aggregates can be inverted~\cite{dinur2003revealing}, and sparse records re-identified from side data~\cite{narayanan2008robust}. Re-pseudonymizing smart meter data does not stop profiling attacks~\cite{cretu2024re}; yearly code changes in masked offer data fail for the same reason, since a unit's identity is carried by its capacity and operating pattern and not by its code. Our two-layer release follows work on the bias of non-negative mechanisms~\cite{mcglinchey2020nonnegative,zhu2021bias}, consistent post-processing~\cite{lee2015maximum}, and the Census noisy measurement files~\cite{abowd2022census}.
{}
\section{Conclusion}
\label{sec:conclusion}

Masking is not a middle ground between transparency and confidentiality. Masking names does not reliably hide generators. With public EIA and EPA data alone, an attacker places \R{reid-units-2025} of ERCOT's labeled masked units at their plants in 2025 and \R{reid-units-2026} in 2026, and puts the right plant among its top five guesses for \R{reid-top5-2025} and \R{reid-top5-2026}. These are not most units, but a release meant to keep every unit's offers confidential fails each unit it names, and the owner of a unit cannot rely on being among the rest. Because PJM's feed carries the same fields and its fleet is about as distinctive, we expect masking to fail in the same way there. Exact aggregates do not hide offers either, since an attacker who knows or models the other units in a total can subtract them. A monthly release by fuel group and price bin, with Gaussian noise calibrated so that each unit's full year of offers is protected, keeps the market supply curve, the clearing price and the capacity of the large fleets at $\eps=10$, and the ratio of a quantity's true spread to the noise predicts which other uses survive. What such a release cannot keep, hourly and unit-level detail, belongs with the market monitor, who already has the unmasked data.

We expect the same weakness wherever a release masks the identity of a physical asset while a sibling public release records the same asset's attributes under its name. The re-identification attack uses nothing specific to generator offers. It needs a masked record whose fields reflect physical properties, such as a capacity, a minimum output or the days on which an asset ran, and a public dataset that lists the same properties by name, as EIA-860 and CEMS do for generators. ERCOT's 60-day disclosure contains other tables of this kind. Generators' ancillary-service offers come from the same resources, so a release that masked them with the same codes would inherit every match made on the energy offers, and energy storage resources offer power limits that EIA-860 also lists for each storage facility, together with its energy capacity. Large industrial loads that sell interruptible service are a weaker case, because no national registry lists them as EIA-860 lists generators. Bids that express a participant's financial positions rather than an asset's physical limits have no such sibling release, and we do not expect the attack to transfer to them. Exact totals, by contrast, can be differenced whatever they count, and the private release of Section~\ref{sec:design} applies to any table whose records can be capped per subject and summed. We have not tested these cases, and we offer them as the directions in which our measurements suggest the problem extends.
{}

\FloatBarrier
\bibliographystyle{ACM-Reference-Format}

\appendix
\section{Proofs}
\label{app:proofs}

\begin{proof}[Proof of Lemma~\ref{lem:sensitivity}]
A unit belongs to one group, so only $X_{m,g}$ for its group changes, by $\frac{1}{|T_m|}\sum_t x_{u,t}$. By the triangle inequality its $\ell_2$ norm is at most $\max_t \|x_{u,t}\|_2 \le \max_t \|x_{u,t}\|_1 \le C$. The same argument with $\|s_{u,t}\|_2\le\|s_{u,t}\|_1 \le 1$ bounds $S_m$.
\end{proof}

\begin{proof}[Proof of Theorem~\ref{thm:privacy}]
Each of the 24 Gaussian releases satisfies $\rho/24$-zCDP by the lemma and the Gaussian mechanism~\cite{bun2016concentrated}. zCDP composes additively, converts to $(\eps,\delta)$-DP as stated~\cite{bun2016concentrated}, and is closed under post-processing.
\end{proof}

\begin{proof}[Proof of Proposition~\ref{prop:trunc}]
Take one cell with value $a\ge0$ on one dataset and $a'\ge0$ on its neighbour, $d=a-a'$, and noise scale $\sigma$. The conditioned release has density $p_a(y)=\varphi((y-a)/\sigma)/(\sigma\Phi(a/\sigma))$ on $y\ge0$. Completing the square in $p_a^{\alpha}p_{a'}^{1-\alpha}$ gives, for $\alpha>1$ and $m=a+(\alpha-1)d$,
\[
  (\alpha-1)D_\alpha(p_a\|p_{a'}) = \frac{\alpha(\alpha-1)d^2}{2\sigma^2} + \ln\frac{\Phi(m/\sigma)}{\Phi(a/\sigma)} + (\alpha-1)\ln\frac{\Phi(a'/\sigma)}{\Phi(a/\sigma)}.
\]
The function $\ln\Phi$ is increasing and concave with slope $\sqrt{2/\pi}$ at 0, so raising its argument by $t\ge0$ from a point $\ge0$ raises it by at most $\sqrt{2/\pi}\,t$. If $d\ge0$, the last term is at most 0 and the middle term at most $(\alpha-1)\sqrt{2/\pi}\,d/\sigma$. If $d<0$, the middle term is at most 0 and the last term at most $(\alpha-1)\sqrt{2/\pi}\,|d|/\sigma$. Either way
\[
  D_\alpha(p_a\|p_{a'}) \le \frac{\alpha d^2}{2\sigma^2} + \sqrt{2/\pi}\,\frac{|d|}{\sigma}.
\]
The cells are independent, so the bounds add over cells. One unit changes a month's $X_m$ by at most $C$ in $\ell_2$ (Lemma~\ref{lem:sensitivity}) and in $\ell_1$, since the cap bounds $\|x_{u,t}\|_1$; it changes $S_m$ by at most 1 in both norms. Each of the 24 releases therefore has $D_\alpha \le \alpha\rho/24 + \sqrt{2/\pi}\,C/\sigma_X$, using $C/\sigma_X = 1/\sigma_S = \sqrt{\rho/12}$. Rényi divergences add under composition, which gives $(\xi,\rho)$-zCDP. The conversion to $(\eps,\delta)$-DP is that of~\cite{bun2016concentrated}: $\eps = \min_{\alpha>1}\,\xi+\alpha\rho+\ln(1/\delta)/(\alpha-1) = \xi+\rho+2\sqrt{\rho\ln(1/\delta)}$.
\end{proof}
{}
\section{Limitations and ethics}
\label{sec:limits}

\paragraph{Limitations.} We study one market over \R{use-months} months. ERCOT has combined cycle at the margin in every month, which makes the marginal fuel trivial to recover and leaves the marginal emissions rate little true variation to track, so these uses may fare differently in markets with a more varied margin. Our conclusion about PJM rests on the two conditions measured in Section~\ref{sec:pjm}, not on an attack scored against PJM's units, and PJM's extra fields could make the attack stronger there in ways we have not tested. $\eps=10$ per unit-year is a policy choice for which we report results, not a value we recommend as safe. The 600~MW cap alone leaves the supply curve \R{curve-cap-only} off. Our attacks are strong but not optimal, so their success on today's releases is a lower bound on what an attacker could learn, while for the private release the DP guarantee bounds every attack, including better ones. The plant attack sums units into daily totals while the plant-year release caps hourly ones. Our 2026 data end in September and rest on EIA's partial monthly sample, so the 2026 re-identification rates understate what an attacker with the full year's public data would reach (Section~\ref{sec:reid}). Finally, our labels are audited but cover only units whose names identify a plant, so re-identification rates are measured over that subset.

\paragraph{Ethics.} We use public data only. ERCOT publishes unit names, so the re-identification reveals nothing ERCOT does not already publish. We built the masked release ourselves from PJM's published rules (Section~\ref{sec:pjm-masking}) and did not run any attack on PJM's data. The paper and our code contain no unit names or unit-to-plant matches. The one exception is the example of Figure~\ref{fig:example}, which shows a plant's idle days from public CEMS data and so could be matched to that plant, but the unit it belongs to appears under its own name in ERCOT's release. Market monitors keep access to unmasked data, so a private public release costs oversight nothing.
{}

\end{document}